\RequirePackage{tikz}
\documentclass[pdflatex,sn-aps,iicol,a4paper]{sn-jnl}

\usepackage{caption}
\usepackage{float}
\usepackage{subcaption}
\usepackage{commath}
\usepackage[T1]{fontenc}
\usepackage[utf8]{inputenc}
\usepackage{xspace}
\usepackage{textpos}

\def\pythia{\textsc{Pythia}8\xspace}
\def\pyt{\textsc{Pythia}\xspace}
\def\mrm#1{\mathrm{#1}}

\def\ee{\ensuremath{\mrm{e}^+\mrm{e}^-}\xspace}
\def\pp{\ensuremath{\mrm{pp}}\xspace}
\def\qb{\ensuremath{\bar{q}}\xspace}
\def\qqb{\ensuremath{q\qb}\xspace}
\def\pap{\ensuremath{\mrm{p}\bar{\mrm{p}}}\xspace}
\def\pL{\ensuremath{\mrm{p}\Lambda}\xspace}
\def\paL{\ensuremath{\mrm{p}\bar{\Lambda}}\xspace}
\def\apap{\ensuremath{\bar{\mrm{p}}\bar{\mrm{p}}}\xspace}
\def\apL{\ensuremath{\bar{\mrm{p}}\Lambda}\xspace}
\def\apaL{\ensuremath{\bar{\mrm{p}}\bar{\Lambda}}\xspace}

\def\pA{\ensuremath{\mrm{p}A}\xspace}
\def\AA{\ensuremath{AA}\xspace}

\def\eg{\emph{e.g.}\xspace}

\def\qcdcr{QCDCR\xspace}

\jyear{2023}%

\begin{document}
\title[~]{Baryon correlations in \pyt}

\author[]{\fnm{Leif} \sur{L{\"o}nnblad}}\email{leif.lonnblad@hep.lu.se}

\author[]{\fnm{Harsh} \sur{Shah}}\email{harsh.shah@hep.lu.se}

\affil[]{\orgdiv{Dept.\ of Physics}, \orgname{Lund University}, \orgaddress{\street{S{\"o}lvegatan\ 14A}, \city{Lund}, \postcode{SE-223~62}, \country{Sweden}}}

\abstract{We present the results from our investigation of angular correlations between baryons pairs in the \pythia event generator. We show how colour reconnection models and hadronization mechanisms influence such angular correlations and in particular address the effect of gluons on the baryon production mechanism in the Lund string fragmentation model. We conclude by discussing the new theoretical ideas in comparison with the ALICE $\pp$ collision results for the baryon angular correlations. We propose a hypothesis for suppressing baryons produced in gluon jets and show how that may influence the angular correlations.}





\maketitle

\section{Introduction}
\label{S:intro}

One of the research interests in particle physics is understanding the
production mechanism and the spatial distribution of particles produced in
high-energy particle collisions. This can be studied in various ways
in colliders, \eg, by measuring single particle distributions as a
function of one or more variables or looking at correlations between
particles, for different species of particles. Using phenomenological
models we can then use these measurements to gain theoretical insights
into the underlying particle production mechanisms.

In this work, we address a long-standing open question about the
angular correlations of pairs of the produced hadrons. A two-particle
correlation function provides information regarding the production of
another particle near the first particle. It is
often studied as a function of
relative pseudorapidity ($\Delta\eta$) and azimuthal angle
($\Delta\phi$) between two particles.

Depending on the chosen range of $\Delta\eta$, the angular
distribution can be studied for long-range (large $\Delta\eta$) or
short-range ($\Delta\eta \sim 0$).  The long-range correlations around
$\Delta\phi \sim 0$ are known as the near-side "ridge". They are studied
extensively in different collision systems like \pp, \pA, and \AA to
understand the collective behaviour of the produced particles (see,
\eg, \cite{CMS:2010ifv} and references therein). The correlation
function is defined such that it is unity for completely uncorrelated
pairs of particles, and any correlation will show up as a larger
value, while lower values indicate that there is an anti-correlation.

The short-range ($\mid\Delta\eta\mid < 1.3$) two-particle angular
correlations were studied by the ALICE experiment in
\cite{ALICE:barcorr14, ALICE:barcorr16} for low transverse momentum
($p_\perp < 2.5$ $\mathrm{GeV}$) hadrons produced in \pp collisions at
$\sqrt{s} =$ 7 $\mathrm{GeV}$.  This angular correlation study shows
that the identified hadron pairs have different angular distributions
depending on the types of hadrons in the pairs. The meson pairs of the
same-sign and opposite-sign particles show correlations peak near
$\Delta\phi = 0$ and a wide bump near $\Delta\phi = \pi$ (also known
as the jet peak and the away-side ridge). On the other hand, baryons
behave differently whether the angular distributions are produced for
the same-sign or opposite-sign baryon pairs. For the opposite-sign
baryon pairs the angular distribution is similar to that of the meson
pairs, with a visible peak near $\Delta\phi = 0$ and almost flat
distribution around $\Delta\phi = \pi$. For the same-sign baryon
pairs, however, there is a clear anti-correlations near
$\Delta\phi = 0$ (except for an indication of a tiny peak for
$\Delta\phi=0$), and a broad peak is observed around
$\Delta\phi = \pi$.

When comparing the ALICE experiment results with \pythia
\cite{Bierlich:2022pfr} generated events, the angular correlations for
the same- and opposite-sign meson pairs are well reproduced, but
\pyt is not able to reproduce the angular correlations for any of the
baryon pairs types. It is also observed that this peculiar behaviour
in the baryon sector is independent of the flavours of the baryons in
the pairs, hence ruling out that the Fermi-Dirac correlation effects
could play a major role. Some suggestions and hypothesises are
proposed in \cite{ALICE:barcorr16}. Following these suggestions,
recently \pyt's hadronization mechanism was studied by a theory group
\cite{Demazure:2022gfl}.

It can be noted that one of the heavy-ion collision experiments, STAR,
measures anti-correlations around $\Delta\phi = 0$ for \pap
pairs produced in Au-Au collisions \cite{STAR:barcorrAuAu}. These
results show that, unlike the observed correlations in \pp collisions,
anti-correlations are observed for \pap pairs in heavy-ion
collisions. Furthermore, if we look into \ee collisions, then \pyt is
able to reproduce the baryon angular correlations
\cite{TPCTwoGamma:barcorree} in \ee collisions. These results from
different collision systems reflect the non-triviality of the
underlying physics of the angular correlations in the baryon sector.
Hence we have decided to investigate the discrepancy in the angular
correlations for baryons produced in \ee collisions and in \pp
collisions. Moreover, we want to identify if any of the event
simulation stages have any significant role in the baryon angular
correlations.

Phenomenological models like \pyt play important roles in our attempts
to quantify the initial and final state effects on the observables.
\pyt is one of the successful general-purpose event generators, which
can reproduce a variety of observables in good agreement with the data
for \ee and \pp collisions for a wide range of collision energies.
The partons are produced during the hard scattering, multiple partons
interactions (MPIs) \cite{Sjostrand:1987su}, and the parton showers, in
stages in \pyt. These produced partons are then treated in terms of
chains of colour dipoles between them, forming strings. An important
feature in hadron collisions is that colour connections between the
partons can be re-arranged by a colour reconnection (CR)
\cite{Sjostrand:1987su, QCDCR} model. After the CR, the colour singlet
strings are hadronized by the Lund string
fragmentation model \cite{LSM} in \pyt. All these
steps can influence the production rate of different hadrons, and
correlations of the hadrons in the simulation results.

For simplicity, we keep our investigation limited to \pp collisions in
this paper. We first have to understand which new effects appear in
the event simulation when we move from \ee collisions to \pp
collisions. Since \pythia is able to reproduce the angular
correlations for the same- and opposite-sign meson pairs fairly well,
we do not discuss the mesons' angular correlations in the rest of the
paper. Instead, our investigation is focused on the angular
correlations of the same- and opposite-sign baryon pairs. We also keep
our results limited to protons while discussing various theoretical
aspects.

In the following, we will start in section \ref{sec:popcorn} by
outlining the main baryon production mechanism in the \pythia
implementation of the Lund string fragmentation model. Special
attention is given to the role of gluons and how they may affect the
production of baryons. This is followed by section \ref{sec:junctions}
where we discuss an alternative way of obtaining baryons in \pyt using
the QCD-inspired colour reconnection model. In
section~\ref{sec:final-state-effects} we then discuss final-state
effects and how they could affect baryon correlations, with special
attention to the hadronic rescattering model. In section
\ref{sec:pheno} we then look at the phenomenology of these models and
try to understand better what could cause the anti-correlation between
like-sign baryons as found in data. Finally, in section \ref{sec:DS}
we summarise with a discussion and an outlook.

\section{Baryons, popcorn and gluons in the Lund Model}
\label{sec:popcorn}

Throughout the perturbative phase of the generation of an event in
\pyt, from multiple scatterings, initial- and final-state showers, the
tracing of colour connections between partons is done using a
leading-colour ($N_C\to\infty$) approximation. In hadronic collisions
there is a possibility to rearrange these connections, as described
below in section \ref{sec:junctions}, but the end results is in any case
in colour-singlet \emph{strings}, each connecting an anti-quark with a
quark via a chain of colour-connected gluons. In the Lund string
model, these strings are fragmented into hadrons as the string
breaks by quark--anti-quark production in the string-like colour field
between the partons.

The production rate of different hadron species depends on their quark
content, mass, and spin. The quarks and anti-quarks of different
flavours are produced in accordance with various parameters in the
Lund String Fragmentation mechanism. The values of these parameters
are primarily fixed from the model comparisons with LEP data. In a
string breaking, a quark and an anti-quark are produced as virtual
particles, which can come on-shell using the energy stored in the
string through a tunnelling mechanism. Clearly, the production of
heavier quarks would then need more of the string energy than light
ones and are therefore suppressed.

The sequence of the further string break-ups will decide if the string
piece will form a meson or a baryon as a primary hadron. A series of
string breaks of multiple \qqb pairs will produce mesons. The simplest
model for baryon production assumes that the string may break by the
production of a diquark-antidiquark pair. This we call the 'diquark
model' in \pyt \cite{Bar:diQ}. The consecutive string breaks of \qqb
pairs on either side of the diquark and anti-diquark will form a baryon
and an anti-baryon.

In the diquark model, the baryon and anti-baryon are always produced
next to each other in rank, and therefore close in rapidity.
Experimental results show that this is not the case always
\cite{TPCTwoGamma:1985zxy}. A mechanism was developed to add
separation between a baryon and an anti-baryon produced next to each
other in the same string. It is called a $\textit{popcorn mechanism}$
\cite{Bar:popcorn}, and adds a possibility of meson
production between the baryon and anti-baryon pair. The idea of the
popcorn mechanism for baryon production is favoured by the
experimental results \cite{TPCTwoGamma:1985zxy}. At the moment, the
popcorn mechanism is enabled by default although only one meson is
allowed to form between the baryon and anti-baryon pair in \pythia.

With or without popcorn, it is clear that we expect some correlations
between baryons and anti-baryons. In particular, if we consider the
case where they are produced next to each other along the string,
their diquark and anti-diquark will have opposite transverse momentum
along the string giving an anti-correlation in azimuth angle. However,
there is no clear way of obtaining baryon-baryon correlation in the
string fragmentation model as such. In a string, there must be at least 
one anti-baryon between two baryons, and the way
transverse momentum is treated in the Lund model, there should be
no correlation between them at all.

The MPI machinery for hadronic collisions produces many strings in an
event, but they are hadronized independently and would not give rise
to correlations between baryons from different strings. It has,
however, been shown that the colour reconnection model in \pythia
gives rise to radial flow \cite{collectivity}, which in principle
could be responsible for the correlations, and we will discuss that in
section \ref{sec:junctions}. Irrespective of colour reconnection it is clear
that the strings in hadronic collisions in general are connected to
partons from MPI scatterings and are therefore not parallel
to the beam axis.

\begin{figure*}
  \begin{center}
  \begin{subfigure}{.5\textwidth}
    \includegraphics[width=\linewidth]{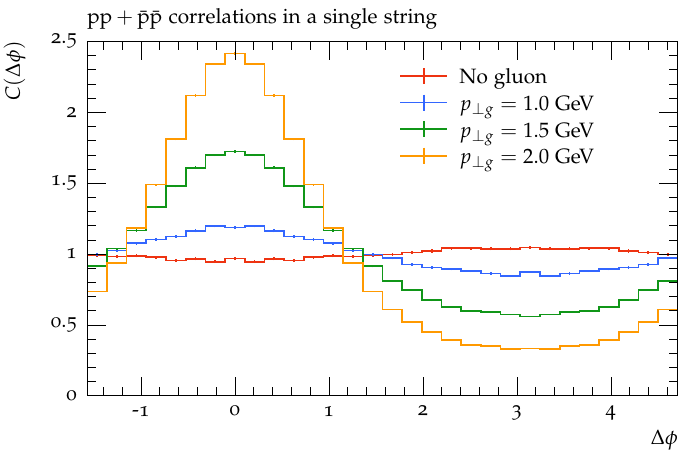}
    \end{subfigure}%
    \begin{subfigure}{.5\textwidth}
    \includegraphics[width=\linewidth]{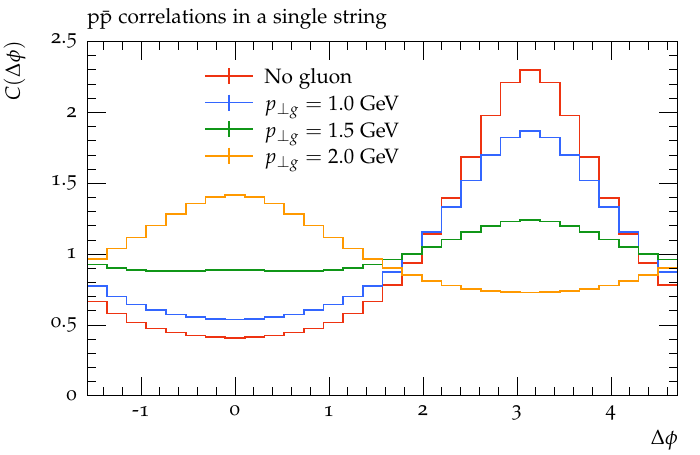}
    \end{subfigure}
  \end{center}
  \caption{Azimuthal correlations along a string with or without
    a soft ($p_\perp=1.0$, 1.5, and 2.0~GeV) gluon. The left plot shows
    proton-proton ($\pp+\apap$) correlations, while the right shows
    proton--anti-proton correlations. The string is spanned between a
    quark and an anti-quark with opposite momenta ($p_{q/\bar{q}}=\pm100$~GeV)
    along the $z$-axis and the gluons are placed at $\eta=0$. Only
    protons with $\mid\eta\mid<1$ are considered.}
\label{fig:single-string}
\end{figure*}

In figure \ref{fig:single-string} we show how a jet peak evolves by
comparing baryon azimuthal correlations in a single straight string,
parallel to the beam axis, with the situation where this string has a
(soft) gluon inserted, giving a transverse ``kink''. For same-sign
protons, the straight string has almost no correlations, but already a
gluon with $p_\perp=1$~GeV will give a rather strong correlation. It
can be noted that in \pythia, around 80\% of all hadrons in the
central pseudo-rapidity bin come from string pieces connected to a
parton with a $p_\perp$ of more than 1~GeV for a 7~TeV \pp\
collision. For \pap we see, as expected, a strong anti-correlation
since the di-quark breakup gives opposite transverse momenta for the
baryon and anti-baryon. But we see that with a soft gluon, the
anti-correlation reduced, and for a 2~GeV gluon it has been turned
into a rather strong correlation.

In the MPI machinery, the string with a gluon would be accompanied by
another string connected to a gluon going in the opposite azimuth
direction. The latter would not be strongly correlated in rapidity,
but would give rise to the so-called away-side ridge in a two-particle
correlation spectra.

\subsection{Gluons vs.\ popcorn}
\label{sec:gluons-vs-popcorn}

Since we now have shown that gluon (mini-) jets contribute to the
baryon angular correlations, it is relevant to scrutinise the
baryon production in a Lund string with gluon ``kinks'' a bit closer.

\begin{figure}
  \centering
  \includegraphics[width=\linewidth]{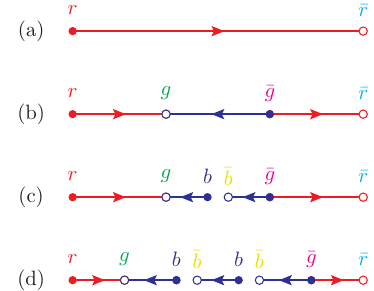}
  \caption{Illustration of the popcorn mechanism. In (a) no
    fluctuation has occurred, and a full string is spanned between a
    red--antired \qqb pair. In (b) a green--antigreen pair has
    appeared on the string as a quantum fluctuation. If the red and
    green quarks form an antiblue triplet, this reverses the colour
    flow in this part of the string, and the net force acting on the
    green quark is zero. In (c) the string breaks by the production of
    a blue--antiblue \qqb pair, resulting in two string pieces with
    diquark ends. In (d) another breakup in the blue triplet
    field results in an additional meson.}
  \label{fig:popcorn}
\end{figure}

The general idea behind the popcorn model used in \pyt is that the
creation of a virtual \qqb pair in a string does not necessarily break
the string. To do that it has to have the right colours such that the
string is divided into two colour singlets. If the colour of the
virtual pair does not match the colours of the string ends, the
virtual fluctuation can then live for a while before the pair is
annihilated again. As an example, in figure \ref{fig:popcorn} we
consider a string stretched between a red quark and an anti-red
anti-quark, then imagine a virtual green--anti-green \qqb pair being
created where the quark is moving towards the red end, and vice
versa. The field between the virtual quarks will then effectively
become antiblue--blue, and if another virtual pair occurs in
this region the string can break. With the two quarks moving towards
the quark end and two anti-quarks moving towards the anti-quark
end. We created two string pieces, each carrying a non-zero
baryon number.

For the \qqb-fluctuation to live long enough for the string
to break in between, the momenta of the $q$ must be longitudinal
towards the quark end of the string and vice versa for the \qb. Any
transverse momenta ($k_\perp$) would be suppressed with a factor
$\propto\exp(-\pi k_\perp^2/\kappa)$, where $\kappa$ is the string
tension. Also, if the $q$ and \qb had opposite momenta, the field in
between would effectively be between two octet charges, which have more
than twice the string tension\footnote{see, \eg,
  \cite{Bierlich:2014xba}.} giving rise to an extra attractive force
between the virtual \qqb pair, making long-lived fluctuations heavily
suppressed.

\begin{figure*}
\centering
\begin{tikzpicture}

@Draw colour strings
\draw (-7.8,-3) -- (-6,-0.6) -- (-3,-0.6) -- (-1.2,-3);
\draw (-5.8,-3) -- (-4.5, -1.5) -- (-3.2, -3);
\draw [dashed] (-8,-3.1) -- (-1, -3.1);

@Name the nodes
\node at (-1.5, -2) {\large$C$};
\node at (-4.5, 0) {\large$B$};
\node at (-7.5, -2) {\large$A$};

@Strings
\draw [green] (-5.8,-3) circle (2pt) node[right]{$\bar{q}$};
\filldraw [green!50!blue] (-4.5, -1.5) circle (2pt) node[below]{$g$};
\filldraw [blue] (-3.2, -3) circle (2pt) node[left]{$q$};
@arrows
\draw[->] (-5.9,-3) -- (-6.2,-3);
\draw[->] (-3.1, -3) -- (-2.8, -3);
\draw[->] (-4.5, -1.5) -- (-4.5, -1.2);

@arrows outer-lines
\draw[->] (-5,-0.6) -- (-5.3,-0.6);
\draw[->] (-4,-0.6) -- (-3.7,-0.6);
\filldraw [green!50!blue] (-4.5, -0.6) circle (2pt) node[below]{$g$};
\draw[->] (-7.9,-3) -- (-8.1,-3);
\draw [green] (-7.8,-3) circle (2pt) node[right]{$\bar{q}$};
\draw[->] (-1.1,-3) -- (-0.9,-3);
\filldraw [blue] (-1.2, -3) circle (2pt) node[left]{$q$};

@popcorn virtual quark pair
\draw[blue] (-3.5, -0.6) circle (2pt) node [above] {$\bar{q}$};
\filldraw[blue] (-2.5, -1.3) circle (2pt) node [right] {$q$};
\draw [blue] (-3.2, -0.6) -- (-3,-0.6) -- (-2.5, -1.3);

\end{tikzpicture}
\caption{A schematic diagram shows two different phases of the
  movement of a $\Bar{q}gq$ string, where the initial momentum of the
  $q$(\qb) is along the (negative) $z$-axis while the gluon momenta is
  perpendicular to them. The innermost lines represent the initial
  phase where a quark and an anti-quark are connected with a gluon
  kink in between. As the string stretches out and moves, the gluon
  gradually loses its energy to the string and eventually stops. At
  this point, the string cannot move further upwards, and the gluon
  kink is basically split into two kinks, and we enter the phase shown
  with the outermost lines. We thus end up with three pieces of
  straight string segments, $\textbf{A}$, $\textbf{B}$, and $\textbf{C}$.
  A virtual popcorn $q\Bar{q}$ pair creation similar to figure \ref{fig:popcorn} (b), but across a gluon kink between the string segments $\textbf{B}$ and $\textbf{C}$ is shown to highlight how the $\qqb$ fluctuation has to propagate across a gluon kink with non-longitudinal momenta.}
\label{fig:gk_area}
\end{figure*}
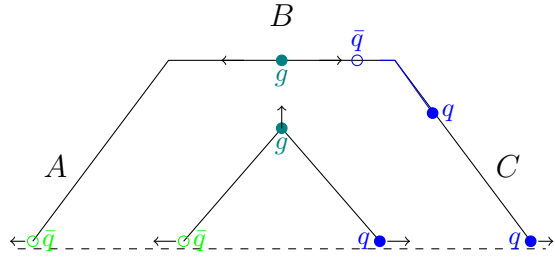

This picture works nicely for a straight string. But if there is a gluon
along a string, the picture changes. In figure \ref{fig:gk_area}
we show two snapshots of a string stretched between a quark and an
anti-quark via a gluon. At first, there are two straight-string
pieces (A and C) with the gluon as a kink. But the gluon is here
retarded by two string pieces and will eventually stop, resulting
in a new straight string piece (B) being formed, and the gluon
kink is split into two.
In figure \ref{fig:gk_area} we show a virtual $q\Bar{q}$ fluctuation across a gluon kink, and
we can notice that $q$ and $\qb$ have non-longitudinal momenta components because they have to propagate to two different string segments (B and C).

In the current implementation of string fragmentation in \pythia,
there is no special treatment of baryon production close to such gluon
kinks. From the description of the popcorn model above, however, it is
clear that for a non-breaking virtual \qqb fluctuation, it would be
very difficult for the quark or the anti-quark to propagate across
such a kink. The pair should have only longitudinal momenta along the
string piece where they are created, but the propagation across
the kink corresponds to non-zero transverse momentum in the string piece
on the other side of the kink, such fluctuations would be suppressed.

Since we have here shown that gluons are important for the azimuthal
correlations between baryons we will in section \ref{sec:suppr-bary-proc} 
use a toy model to investigate the
possible effects of the suppression of baryon production close to
gluon kinks.

\section{Junctions and colour reconnections}
\label{sec:junctions}

The popcorn and di-quark models are not the only way of obtaining
baryons in \pyt. In some cases non-trivial colour topologies may arise
prior to the string fragmentation stage, \eg, from the treatment of
remnants in hadronic collisions, or when looking at baryon number
violating BSM processes. In the MPI machinery, it is not uncommon that
two (valence) quarks are taken from a proton, leaving a remnant in a
colour-triplet state. Similarly, baryon number violating processes may
decay a colour-triplet particle into two anti-triplet particles. In
both cases, we may obtain colour-singlet string systems connecting
three quarks (or three anti-quarks) in a so-called string junction
topology \cite{Sjostrand:2002ip}. \pythia is able to hadronise such
systems, in a process that always will produce a net baryon number.

We will not be concerned with BSM here, and the junctions formed in
the MPI remnant treatment mainly affects baryons in the far forward or
backward regions of rapidity. There is, however, another way of
creating junctions available in \pythia, using the so-called QCD
colour reconnection model \cite{QCDCR}.

CR models re-arrange the colour connections of the colour dipoles
produced after MPIs and parton showers.  The primary objective of CR
is to reduce the net string length so that the model can reproduce the
charged particles' multiplicity and the observed enhancement in
$\langle p_\perp \rangle $ $(N_{ch})$ distribution.
\pyt has a default CR model, which is based on MPIs
\cite{Sjostrand:1987su}, where the different MPIs are colour
reconnected in the $N_{c} \rightarrow \infty$ limit, and the only
criteria to satisfy is to reduce the net string length.

\begin{figure}
  \centering
  \includegraphics[width=\linewidth]{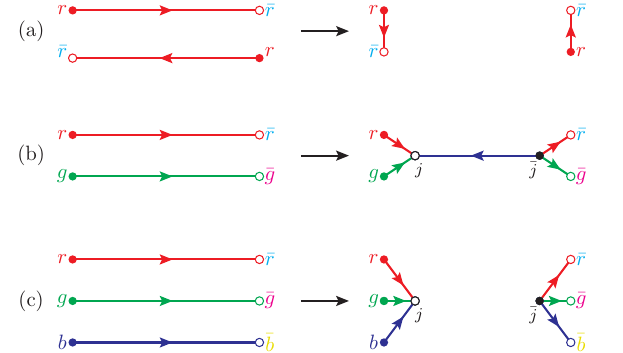}
  \vspace{1mm}
  \caption{Illustration of the possible reconnections in the \qcdcr
    model. (a) A ``swing'' between two dipoles in the same colour
    state. (b) Two dipoles in different colour states can form a
    connected junction--anti-junction system. (c) Three dipoles in
    different colour states form separate junction and
    anti-junction systems. In all cases, the total string length must
    be reduced in the process. Note that the dipole ends may be gluons
    that connect to other dipoles in a string system.}
  \label{fig:colrec}
\end{figure}

\pythia has an alternative model, the \qcdcr model \cite{QCDCR}, which
follows QCD colour rules while performing CR. The \qcdcr model allows
the formation of junction systems, where two or three string pieces
can be colour connected to a junction and an anti-junction system,
each of which will produce at least one (anti-)baryon. The different
colour reconnections possible in this model are summarised in figure
\ref{fig:colrec}. For case (a) two string pieces where the colour end
of one is in a colour-singlet state w.r.t.\ the anti-colour end of the
other, can reconnect in a so-called \textit{swing}. In (b) we instead
have the situation where the two (anti-) colour ends together are in
an anti-triplet (triplet) state and can reconnect to two junction
systems connected by a dipole. Finally, in (c) the (anti-) colour ends
of three dipoles together form a colour singlet and can reconnect into
a separate (anti-) junction system. In each case only reconnections
that reduce the overall string lengths\footnote{The \textit{length} of
  a string is approximately given by the sum of the logarithm of the
  invariant masses of the dipoles forming the string (see
  ref.~\cite{QCDCR} for details).} are allowed. This means that 
  dipoles that are approximately anti-parallel in momentum space are
more likely to reconnect like in case (a), while the opposite is true for
cases (b) and (c).

The number of junctions in \ee collisions is very low, and it is
pointed out in \cite{QCDCR} that the effect of the \qcdcr model is not
clearly visible there. But in \pp collisions there are sometimes many
MPIs, which enhance the possibility of junction formation during the
CR. This means the \qcdcr will produce additional baryons, on top of
what is produced in the subsequent string fragmentation.

It should be noted that the two connected junctions can be separated by a long dipole or a chain of multiple dipoles that are reconnected and can separate the junctions by a large rapidity span.
For example, the junction baryon and anti-baryon produced in figure \ref{fig:colrec} case (b) are often separated by multiple hadrons produced in between due to the fragmentation of the string piece connecting the two junctions.
The junction systems in figure \ref{fig:colrec} case (c) are two independent colour singlet systems, and the baryon and the anti-baryon produced at junctions in these two systems are non-correlated.
Hence the correlation between the resulting baryon and
anti-baryon due to the junctions is much weaker than for the baryon--anti-baryon pairs produced in the string breaking.
We can therefore expect that the correlations will be diluted by the additional baryons from the \qcdcr model.

\section{Final-state effects on correlations}
\label{sec:final-state-effects}

There are many potential final-state effects that may affect
correlations between hadrons produced in the string fragmentation. The
Lund group has studied several such models, e.g., a model for
Fermi-Dirac correlations \cite{DuranDelgado:2007lak}, the so-called
rope hadronisation model \cite{Bierlich:2014xba} and a model for
repulsion between strings \cite{shoving}. Of these, the rope model
mainly affects the flavour composition, and is not expected to give
significant effects on correlations. Also, the string repulsion will
give a flow effect in high multiplicity \pp events, but the effect is
overall quite small in \pp, and it will increase correlations both at
$\Delta\phi=0$ and $\Delta\phi=\pi$ and would therefore not improve the description of
baryon-baryon correlations in \pythia at small angles. Fermi-Dirac
effects would decrease the correlation at small angles for identical
baryons, but again the effect is expected to be small\footnote{We have
  confirmed that the effect is small by making a rudimentary
  implementation in \pythia of the Fermi-Dirac model described in
  \cite{DuranDelgado:2007lak}}. Also, as already pointed out in
\cite{Demazure:2022gfl}, the effect found in \cite{ALICE:barcorr14,
  ALICE:barcorr16} is the same for \pp and \pL, this can also not
improve the situation.

Instead, we will focus on the model for hadronic rescattering
\cite{hadrescat,hardrescat2}. By following the production vertices of
all partons in the event, it is possible to calculate the
production points of all hadrons in the string fragmentation
\cite{Ferreres-Sole:2018vgo}. Then one can study the
possible scatterings between these hadrons in a way similar to the
UrQMD \cite{Bass:1998ca} and SMASH \cite{SMASH:2016zqf}
models. Clearly, the rescatterings will mainly affect hadrons that are
propagating in the same general direction, and one may expect that
it will reduce correlations at $\Delta\phi=0$, and we will therefor
investigate this model in the following section.

\section{Comparison with data}
\label{sec:pheno}

So far we have presented a set of ideas that may affect baryon
correlations in \pyt, and in this section, we will confront these ideas
with data. It can be noted that we have also tested varying standard
string fragmentation parameters, such as flavour ratios, di-quark
production rate, spin ratios of the di-quarks, and $p_\perp$ assignment to
the produced hadrons. We found, however, that none of these changes
significantly affects the angular correlations of the same-sign baryon
pairs in \pyt.

In Ref.\cite{Demazure:2022gfl}, the main conclusion was that only by
forcing \pythia to produce at most one di-quark breakup per string, it
was possible to understand the correlations found in data. Such an
artificial change in the behaviour of the string fragmentation is of
course not a satisfactory solution, but it gives us hints as to what
is needed. We will therefore concentrate on \emph{reducing} the number
of di-quark breakups in strings with (semi-) hard gluons, but also to
introduce alternative baryon production mechanisms that do not
exhibit the correlations found in string fragmentation. In addition, we
will also consider final state effects from hadronic rescattering.

In all simulations, we have used the \pyt version 8.306 to generate
\pp events at $\sqrt{s}=7$~TeV.  The analysis of the generated events
was done using the Rivet \cite{Bierlich:2019rhm} routine
\verb:ALICE_2016_I1507157: which mimics the analysis in
\cite{ALICE:barcorr16}.\footnote{Note that we have made slight
  corrections to the kinematical cuts used for different particle
  species in the Rivet routine, to better reflect the cuts made in the
  experiment. These corrections will be included in a future release
  of Rivet.}

\subsection{The QCD colour reconnection model}
\label{sec:qcdcd-colo-reconn}

\begin{figure*}
\begin{subfigure}{.5\textwidth}
  \includegraphics[width=\linewidth]{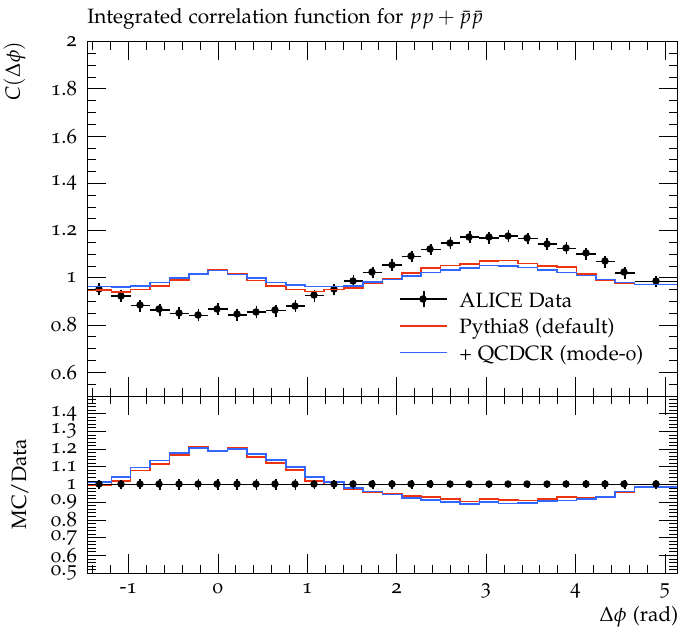}
\end{subfigure}%
\begin{subfigure}{.5\textwidth}
  \includegraphics[width=\linewidth]{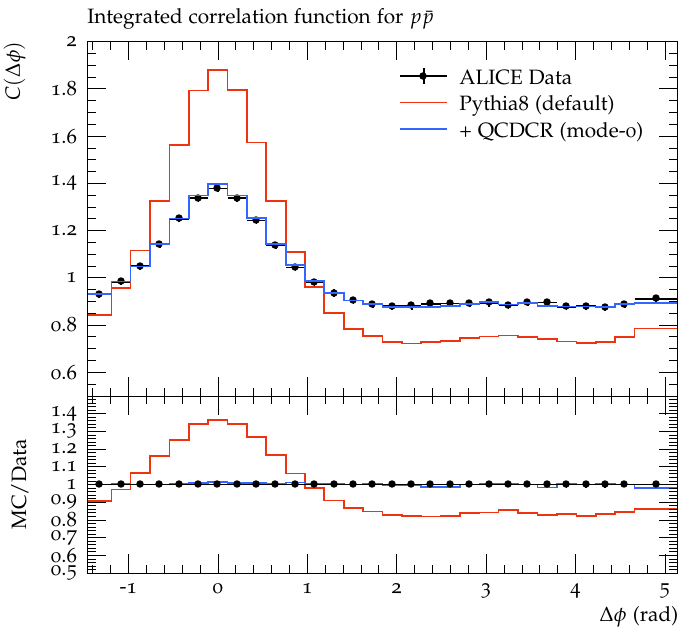}
\end{subfigure}

\begin{subfigure}{.5\textwidth}
  \includegraphics[width=\linewidth]{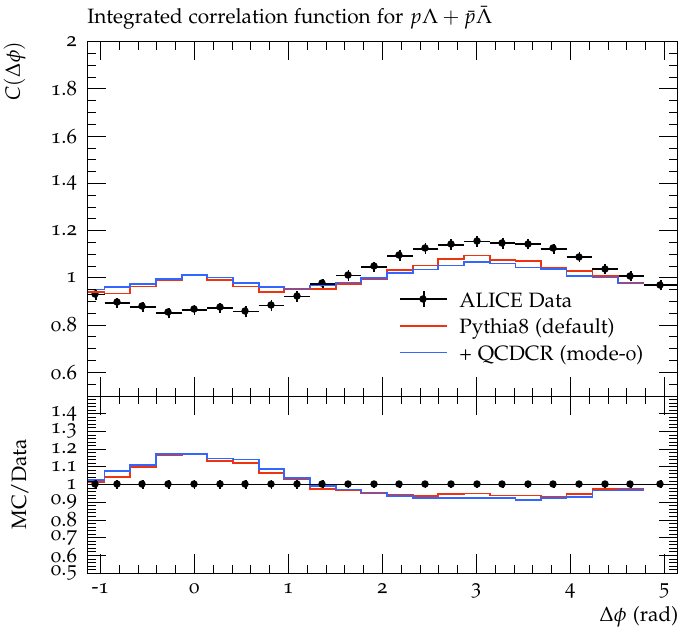}
\end{subfigure}%
\begin{subfigure}{.5\textwidth}
  \includegraphics[width=\linewidth]{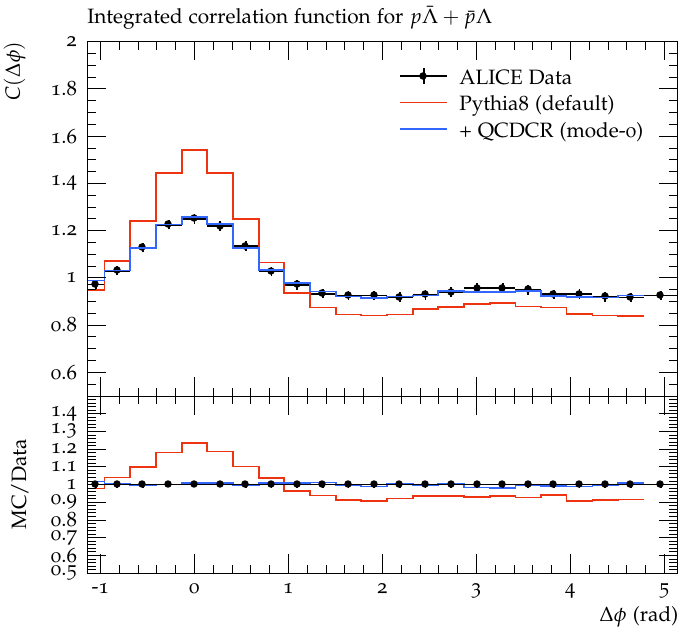}
\end{subfigure}
\caption{Baryon azimuthal correlations. \textit{Top}: \pp + \apap
  pairs on the left, and \pap pairs on the right. \textit{Bottom}: \pL
  + \apaL pairs on the left, and \paL + \apL pairs on the
  right. Events are generated for \pp collisions at $\sqrt{s}=7$~TeV
  and are compared with ALICE data \cite{ALICE:barcorr16}. The red and
  blue lines represent results using the \pythia Monash tune with
  MPIs-based colour reconnection, and using \qcdcr (mode-0) colour
  reconnection respectively.}
\label{fig:pp_CR}
\end{figure*}

We begin with the \qcdcr model, which introduces a completely
new way of producing baryons. We have used the so-called ``mode-0''
tune presented in \cite{QCDCR}, with no further changes. The results
are presented in figure \ref{fig:pp_CR} and show a remarkable
improvement in the baryon--anti-baryon correlations as compared to
the default CR model in \pythia.

The choice of the CR model does not, however, improve the angular
(anti-) correlations for the same-sign baryon pairs. The effect is
rather a general reduction in correlations, which is expected since
the model will produce additional baryons with fewer correlations.

The reduction is more clearly seen for the opposite-sign baryon pairs,
where \qcdcr model reduces the amplitude of the baryon-antibaryon pair
correlations near $\Delta\phi=0$, and also reduces the corresponding
away side anti-correlation, bringing the simulation results in
agreement with the data. It is clear that the separation between the
junction and anti-junction systems created by the \qcdcr model plays a
significant role in improving the angular correlations between the
opposite-sign baryon pairs.

It should be noted that we have also studied the effects in meson
correlations (which are not shown here) but found no significant
effect of the choice of CR model there.

Since the \qcdcr shows significant improvement in the angular
correlations of the opposite-sign baryon pairs, we will in the
following use the \qcdcr as our base-line set-up when adding other
modifications.

\subsection{Hadronic rescattering}
\label{sec:HReSc}

\begin{figure*}
\begin{subfigure}{.5\textwidth}
  \includegraphics[width=\linewidth]{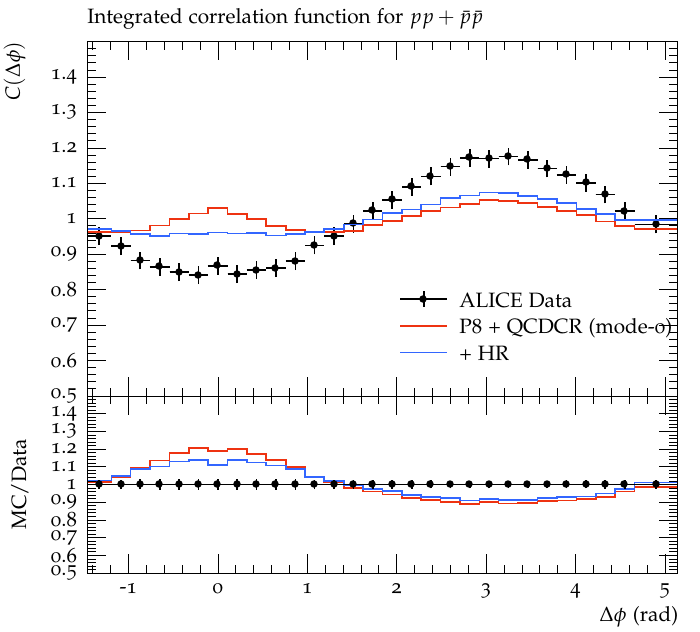}
\end{subfigure}%
\begin{subfigure}{.5\textwidth}
  \includegraphics[width=\linewidth]{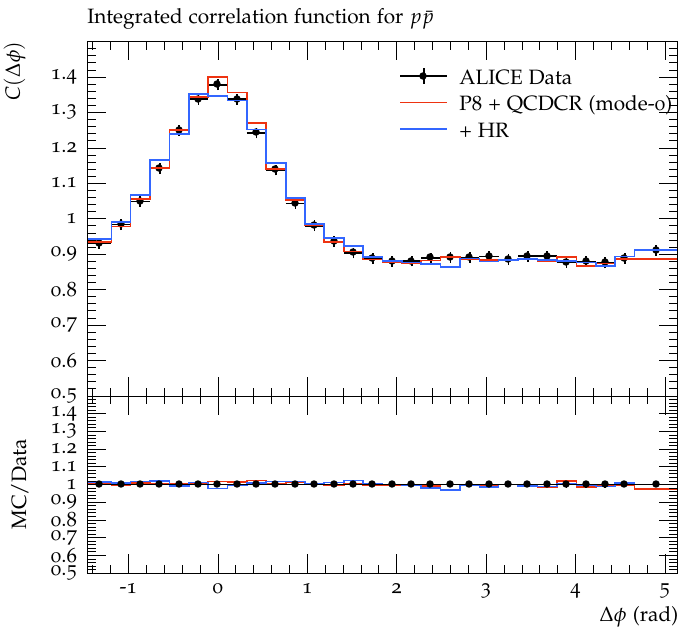}
\end{subfigure}
\caption{Proton azimuthal correlations for \pp + \apap pairs on the
  left, and \pap pairs on the right. Events are generated for \pp
  collisions at $\sqrt{s} = 7$~TeV and are compared with ALICE data
  \cite{ALICE:barcorr16}. The red lines show results from \pythia with
  \qcdcr (mode 0), while blue lines show the same but with hadronic rescattering.}
\label{fig:pp_HReSc}
\end{figure*}

The produced hadrons can interact with
nearby hadrons via elastic or inelastic scattering. A model for
hadronic rescattering \cite{hadrescat} was recently added in \pyt,
implementing $2 \rightarrow 2$ and $2 \rightarrow 3$ type inelastic
and elastic hadronic rescatterings.\footnote{It should be noted that the absence of $3 \rightarrow 2$ scatterings means that the overall multiplicity of an event is slightly increased. In \cite{hadrescat} this was compensated by a slight change in the $p^{Ref}_{T0}$ parameter regulating the divergences in the MPI scattering cross sections. We have checked that a similar increase does not significantly influence the correlation results presented here.}
The naive expectation is that rescattering will blur preexistent
correlations between particles going in the same direction, and that
is indeed what is seen for the \pap\ correlations in figure
\ref{fig:pp_HReSc}. The effect is not very large, but we know that
rescattering effects in general are quite modest in \pp\
collisions and stronger only in high multiplicity events. We note, however, that for the like-sign proton
correlations in figure \ref{fig:pp_HReSc} the effect is much more
visible. In fact, the correlation around $\Delta\phi=0$ is all but
wiped out.

The reason for this is somewhat non-trivial, and is related to the
annihilation of baryons--anti-baryon pairs in the rescattering. As
explained in section \ref{sec:popcorn}, the peak at
$\Delta\phi=\Delta\eta=0$ mainly comes from jets, where the two
particles are typically produced in the same string. For \pap\ pairs
the main contribution is pairs produced in a single diquark breakup,
and since the diquark and anti-diquark will have opposite transverse
momenta along the string, it is very unlikely that the baryons formed
would rescatter with each other. For \pp and \apap pairs, however, we
would need two baryon--antibaryon pairs produced close together along
the string and a baryon in one pair could then more easily annihilate
with the anti-baryon in the other. This effect turns out to be rather
large. Adding rescattering to the $p_{\perp g}=2$~GeV runs in figure
\ref{fig:single-string} does not affect the shape of the correlations
very much, but the number of like-sign pairs will be reduced by around
40\%. We, therefore, conclude that the reason for the relatively large
effect for \pp\ and \apap\ in figure \ref{fig:pp_HReSc} is that the
number of pairs stemming from the same string is reduced.

\subsection{Suppressing baryon production close to gluon kinks}
\label{sec:suppr-bary-proc}

\begin{figure*}
\begin{subfigure}{.5\textwidth}
  \includegraphics[width=\linewidth]{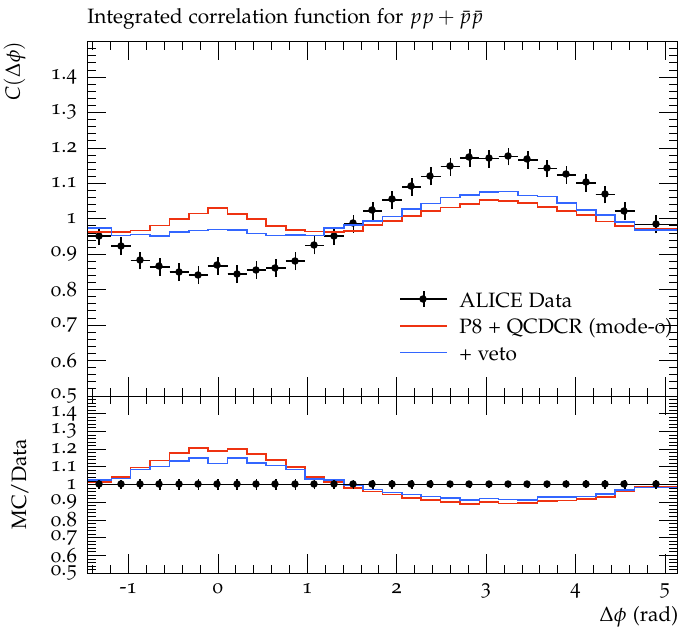}
\end{subfigure}%
\begin{subfigure}{.5\textwidth}
  \includegraphics[width=\linewidth]{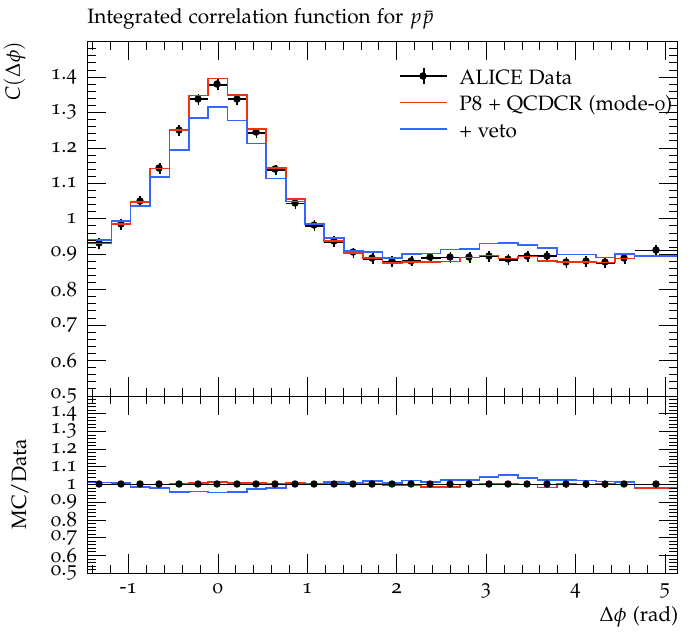}
\end{subfigure}
\caption{Proton azimuthal correlations for \pp + \apap pairs on the
  left, and \pap pairs on the right. Events are generated for \pp
  collisions at $\sqrt{s} = 7$~TeV and are compared with ALICE data
  \cite{ALICE:barcorr16}. The red lines show results from \pythia with
  \qcdcr (mode-0), while blue lines show the same but with a veto
  on primary baryons spanning a gluon kink as explained in the text.}
\label{fig:pp_VtHad}
\end{figure*}

There is currently no proper implementation for the possible suppression
of baryon production in string fragmentation close to gluon kinks in
the popcorn mechanism discussed in
section~\ref{sec:gluons-vs-popcorn}. Instead, we will study a
simplified toy model to understand what the effects may be.

We have decided to constrain the baryon production using the
$\texttt{UserHooks}$ facility in \pythia \cite{Bierlich:2022pfr},
which allows a user to intervene at different stages of the event
generation. In particular, there are options to intervene in the
string fragmentation procedure and one possibility is to simply veto
the production of a single hadron, based on additional criteria
implemented by the user.

In our crude implementation, we veto any baryon produced in a diquark
breakup if the previous breakup was in a different string region. As
an example consider the case in figure \ref{fig:gk_area} where the
gluon has lost all its energy. If there has been a normal \qqb\
breakup in string region \textbf{C} and the next breakup is a
diquark--anti-diquark breakup in region \textbf{B}, we veto the baryon
to be produced, and tell \pyt to try another breakup instead. It
should be noted that the Lund string fragmentation model is
left--right symmetric, and if we go from the other (\qb) end and the
same diquark breakup occurred in region \textbf{B} the following \qqb\
breakup in \textbf{C} producing the same baryon from a \textit{kinky}
string piece, is not vetoed. The reason for implementing it in this
way is technical, but effectively it will result in a suppression of
baryons produced around a string corner with a factor of 0.5.

In figure \ref{fig:pp_VtHad} see the effect of applying this toy model
to \pythia including \qcdcr. As expected the jet peak for baryon pairs
is reduced, and for $\pp+\apap$ the lines are moved closer to
the data. Unfortunately, the reduction of the jet peak is also present for
\pap, which worsens somewhat the excellent agreement with data
obtained from the \qcdcr model.

It should be noted that our toy model will reduce the overall number
of baryons produced in general, even in \ee collisions, since also
there we have gluon kinks. To be completely fair we should therefore
have retuned the parameters affecting baryon production to obtain the
same reproduction of LEP data. We would, however, expect the reduction
of the jet peak to stay more or less the same.

\begin{figure*}
\begin{subfigure}{.5\textwidth}
  \includegraphics[width=\linewidth]{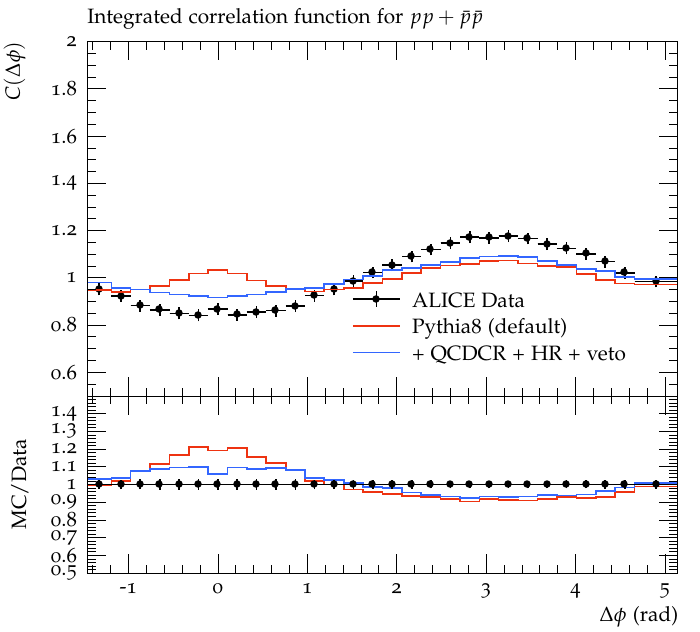}
\end{subfigure}%
\begin{subfigure}{.5\textwidth}
  \includegraphics[width=\linewidth]{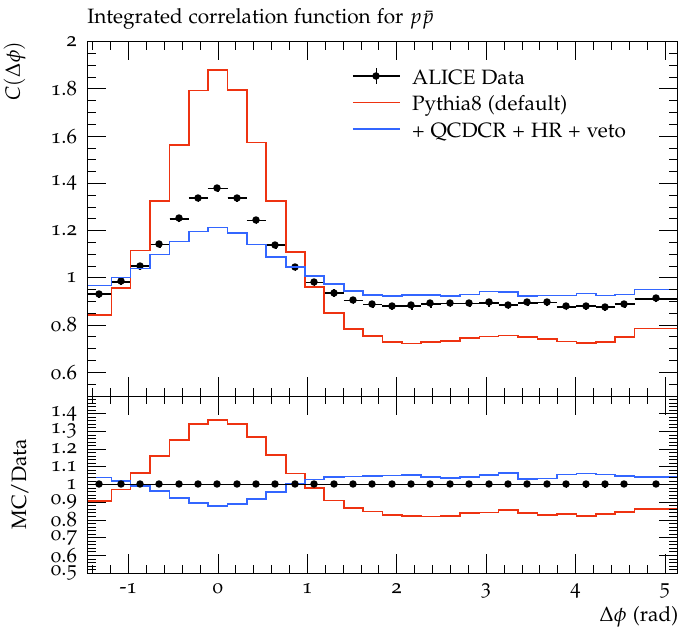}
\end{subfigure}
\caption{Proton azimuthal correlations for \pp + \apap pairs on the
  left, and \pap pairs on the right. Events are generated for \pp
  collisions at $\sqrt{s} = 7$~TeV and are compared with ALICE data
  \cite{ALICE:barcorr16}. The red lines show results from the default
  \pythia, while blue lines show the result with \qcdcr (mode-0) colour reconnection and hadronic rescattering (HR) switched on and with a veto on primary baryons spanning a gluon kink as explained in the text.}
\label{fig:pp_VtHReSc}
\end{figure*}

As a final result, we show in figure \ref{fig:pp_VtHReSc} a comparison
between the default \pyt and the accumulated changes of all three
models investigated her: \qcdcr, hadronic rescattering and the vetoing
of baryons close to gluon kinks. We see that the reproduction of the
ALICE date is far from perfect if the models are added, but there is a
clear improvement over the default \pythia. The jet peak is reduced
both for \pp+\apap and \pap, and we see that there is even an
anti-correlation for like-sign proton pairs around $\Delta\phi=\Delta\eta=0$.

\section{Discussion and summary}
\label{sec:DS}

We have shown that the observed anti-correlation for the same-sign
baryon pairs in the ALICE experiment for \pp collisions is a
non-trivial outcome of many-fold effects. We also presented the first
steps towards understanding the failure of \pyt in reproducing these
experimental results. We found that two already existing models in
\pythia, the \qcdcr model and the hadronic
rescattering model, have a significant effect on the correlations, and adding these to the default \pythia improves the description
of data significantly.

The \qcdcr\ model produces additional baryons due to junction systems
forming as a part of the reconnections of the colour dipoles. Such
junction baryons are much less correlated than those produced in
string fragmentation.
The string connecting two junctions can produce multiple hadrons between the two junction baryons, unlike the popcorn baryons, which are separated by only one meson in between.
We show that it visibly reduces the
correlations between the opposite-sign baryon pairs in the jet peak
near $\Delta\phi=0$. As a result, \pyt is able to reproduce the
angular correlation distribution for the opposite-sign baryon pairs.

The anti-correlations in the same-sign baryon pairs are rather complex
results. Although the \qcdcr model improves the \pyt results, it is
not sufficient. Adding the hadronic scattering model, we found that
the effect of annihilation of baryon--anti-baryon in jets with more
than one baryon--anti-baryon pair is quite significant, while if there
is only one pair, there is typically no annihilation. This gives a
further reduction of the jet peak for same-sign baryons, while the
effects on unlike-sign correlations are small. Still, the jet peak for
same-sign baryons in \pyt needs to be further reduced in order to
reproduce data.

The authors in \cite{Demazure:2022gfl} managed to make \pyt reproduce
data, by forcibly forbidding more than one baryon--anti-baryon pair to
be produced in a string. This effectively removed the jet peak in the
same-sign baryon correlations, leaving only the anti-correlation in
the away-side ridge. Here we instead propose a more physical
mechanism, where baryon production close to gluon kinks in a string is
suppressed. The motivation for this comes from the popcorn model of
baryon production in a string. Here an extra non-breaking virtual \qqb
is required to exist before a \qqb pair breaking occurs to produce
an effective di-quark breakup, and we argue that it is less likely to
have such an extra pair close to a gluon kink.

Since the jet peak around $\Delta\phi=\Delta\eta=0$ in the angular
correlations in \pp collisions mainly consist of particle pairs
from the same (mini-) jet (which at the LHC is likely to be a gluon
jet), one would then expect a reduction of the peak for baryon pairs
in general. For same-sign baryon pairs, we expect the reduction to be
even bigger since we require two such popcorn breakups in the same
string. We have here qualitatively confirmed that this is the case
using a toy model, where we simply disallow some such breakups close
to gluon kinks, which has motivated us to attempt a more realistic
modelling of the effect in the future. Such a model would have to take
into account the size of the transverse momentum of the gluon kink, as
well as the distance between the breakup and the kink. Since the
overall number of baryons would be reduced, such a model would also
require a proper retuning of the baryon parameters in \pythia, but it
is still likely that the jet peak for same-sign baryon correlations
would be reduced. Whether it will be reduced enough to reproduce data
remains to be seen.

We believe that the baryon suppression near gluon kinks will affect the 
baryon-to-meson ratios in \ee collisions as well. Hence the model should be 
retuned to LEP data for \ee collisions in future.

Finally, we note that there are other independent measurements that
could verify our hypothesis of suppressed baryon production close to
gluons. One obvious example is to compare the baryon-to-meson ratio
inside a gluon jet to that of a quark jet, which could be done by
comparing inclusive jets to jets produced together with hard
photons. We are not aware of any study where the jet substructure has
been studied for identified hadrons, but we would certainly like
encourage our experimental colleagues to pursue such measurements at
the LHC.

\bmhead{Acknowledgments}

\label{sec:acknowledgements}

We would like to thank G\"osta Gustafson for coming up with the idea of
suppressing popcorn production close to gluons. We would also like to
thank Christian Bierlich and Torbj\"orn Sj\"ostrand for useful
discussions.

This work was funded in part by the Knut and Alice Wallenberg
Foundation, contract number 2017.0036, Swedish Research Council,
contracts numbers 2016-03291 and 2020-04869, in part by the European
Research Council (ERC) under the European Union’s Horizon 2020
research and innovation programme, grant agreement No. 668679, and in
part by the MCnetITN3 H2020 Marie Curie Initial Training Network,
contract 722104.



\bibliography{articlemain}


\end{document}